\documentclass[sigconf, nonacm]{acmart}

\usepackage{booktabs}
\usepackage{graphicx}

\setcopyright{none}
\renewcommand\footnotetextcopyrightpermission[1]{}
\begin{document}

\title{Securing AI-Generated Code: A Just-in-Time Vulnerability
Detection and Remediation Pipeline}

\author{Mikhail Surikov}
\affiliation{%
  \institution{Georgia Institute of Technology}
  \city{Atlanta}
  \state{Georgia}
  \country{USA}
}
\email{msurikov3@gatech.edu}

\begin{abstract}
AI-assisted development tools generate vulnerable code at significant
rates \cite{wessling2025, ji2024}, yet few automated mechanisms exist to
detect, enrich, fix, and verify security issues at development velocity,
particularly ones that ground remediation in real-world threat context
rather than pattern matching alone. This paper presents an automated
security evaluation pipeline
that generates Python code from LLMSecEval prompts \cite{tony2023}, scans
for vulnerabilities using CodeQL and Bandit in parallel with an
independent Code Validator LLM, enriches the Code Validator findings with
MITRE ATT\&CK techniques, CWE Observed Examples, and Python best practice
guidelines, generates fixes via the Code Generation LLM, and re-scans with
CodeQL and Bandit to verify outcomes. Two pipeline configurations were
evaluated: Pipeline 1 (P1), where the Code Generation LLM receives
enriched Code Validator findings only, and Pipeline 2 (P2), where it
additionally receives the initial CodeQL and Bandit findings. Both
configurations were run across four Claude models: Opus 4.8, Sonnet 4.6,
Sonnet 5, and Haiku 4.5, producing 80 runs against 26 LLMSecEval prompts
covering 9 CWE categories.

P1 reduced static analyzer findings across all four models, ranging from
$-9\%$ (Opus 4.8) to $-54\%$ (Sonnet 5). P2 deepened these reductions
further, ranging from $-29\%$ (Opus 4.8) to $-69\%$ (Haiku 4.5), with P2
outperforming P1 for every model. Verdict consistency averaged
approximately 81\% modal agreement across all configurations, with P2
marginally more stable than P1. Remediation introduced new
vulnerabilities in 15--22\% of cases: roughly 70\% involved a single new
finding, and P2 reduced churn for three of four models, with Sonnet 5 as
the sole exception. Notably, the best Code Generation LLM (Opus 4.8) was
not the best pipeline performer, as Sonnet 4.6 produced the lowest
residual findings and highest pass rate after P2 remediation, suggesting
that pipeline effectiveness and first-draft security are distinct
properties.
\end{abstract}


\maketitle

\section{Introduction}
AI-assisted development tools generate vulnerable code at significant
rates \cite{wessling2025, ji2024}, yet the processes designed to catch
security issues were built for a previous era. Code review, static
analysis in CI/CD pipelines, and security audits all assume human-speed,
human-trusted output. Unlike code from human contributors, AI-generated
code compounds these risks simultaneously: it arrives faster than review
processes can absorb, it carries a documented bias that reduces developer
scrutiny \cite{ji2024}, and it potentially reproduces the same vulnerable
patterns indefinitely, as there is no feedback loop to improve LLMs over
time based on specific examples.

Existing automated tools detect known vulnerability patterns but cannot
reason about the threat context those vulnerabilities enable, neither can
they generate targeted, enriched fixes. Prior work has either measured the
scale of the problem \cite{ji2024, schreiber2025} or demonstrated that
CWE-specific prompting reduces vulnerability rates
\cite{bruni2025, kharma2026}, but grounding remediation in system-level
threat context, specifically MITRE ATT\&CK techniques and CWE Observed
Examples, remains largely unexplored as part of an automated
detection-to-verification loop. This gap is the motivation for this work.

This paper presents an automated security evaluation pipeline that
addresses this gap. In this project, static analysis tools CodeQL and
Bandit are used as a measuring tool for assessing the quality of the
generated code (baseline) and after the pipelines are completed, the
validated and improved code is re-scanned with the same tools, the results
are compared and measured. The pipeline generates Python code from
LLMSecEval prompts \cite{tony2023}, scans with CodeQL and Bandit in
parallel with an independent Code Validator LLM, enriches findings with
MITRE ATT\&CK techniques \cite{mitre2024}, CWE Observed Examples
\cite{mitre2023}, and Python best practice guidelines, generates fixes via
the Code Generation LLM, and re-scans with CodeQL and Bandit to verify
outcomes. Two pipeline configurations are compared: Pipeline 1 (P1), where
the Code Generation LLM receives enriched Code Validator findings only,
and Pipeline 2 (P2), where it additionally receives the initial static
analyzer findings. The evaluation addresses two hypotheses: (H1) does an
enriched Code Validator LLM prompt improve the security of AI-generated
code, and (H2) does adding static analyzer findings to the fix context
further improve remediation outcomes? Both are evaluated across four
Claude models against 26 LLMSecEval prompts \cite{tony2023} covering 9 CWE
categories from the MITRE Top 25: SQL injection (CWE-89), OS command
injection (CWE-78), path traversal (CWE-22), deserialization of untrusted
data (CWE-502), use of hard-coded credentials (CWE-798), cross-site
scripting (CWE-79), improper input validation (CWE-20), insufficiently
protected credentials (CWE-522), and incorrect permission assignment
(CWE-732).

The remainder of this paper is organized as follows. Section 2 reviews
related work. Section 3 describes the methodology. Section 4 details the
implementation. Section 5 presents evaluation results. Section 6 discusses
findings and connects them to prior work. Sections 7 and 8 address
limitations and threats to validity. Section 9 outlines future work.
Section 10 concludes.

\section{Background \& Related Work}

\subsection{Vulnerability rates in AI-generated code}
Recent empirical work establishes that AI coding tools introduce security
vulnerabilities at significant and consistent rates.
Wessling~\cite{wessling2025} found that 45\% of code samples generated by
the AI models failed security tests, and notably that newer and larger
models showed no improvement in security performance. Ji et
al.~\cite{ji2024} identified AI-generated code as a systemic software
supply chain risk, documenting automation bias as a compounding factor:
developers over-trust generated output, skip critical review, and rate
AI-generated code as more secure than human-written code despite evidence
to the contrary. Schreiber and Tippe~\cite{schreiber2025} conducted a
large-scale analysis of 7,703 AI-attributed files in public GitHub
repositories and found that Python exhibits the highest vulnerability
rates across all major AI coding tools, with 12.1\% of files containing at
least one detectable CWE. Taken together, these studies establish that the
problem is real and measurable.

\subsection{Evaluation benchmark}
Tony et al.~\cite{tony2023} introduced LLMSecEval, a dataset of 150
natural language prompts covering 18 CWE categories, each paired with a
manually verified secure reference implementation. LLMSecEval has become
the de facto standard benchmark for evaluating LLM code security and is
used across all major studies in this space, enabling direct comparison of
results. This work uses a subset of 26 prompts covering 9 CWE categories
selected for their relevance to Python and their coverage by the static
analyzers used in the study.

\subsection{CWE-specific prompting as mitigation}
A cross-study consensus has emerged that CWE-specific prompting guidance
consistently outperforms generic security instructions. Bruni et
al.~\cite{bruni2025} systematically benchmarked 20 prompt engineering
techniques across three GPT models and found that a security-aware persona
prefix reduced vulnerability rates by up to 56\%, with an iterative
recursive criticism and improvement (RCI) technique reaching 68.7\%.
Aldosari and Aldawsari~\cite{aldosari2026} demonstrated that CWE-specific
meta-prompting achieves up to 77\% improvement across four programming
languages and three open-source models, also finding that language choice
has a comparable effect on vulnerability rate as model choice. Kharma et
al.~\cite{kharma2026} proposed Mitigation-Aware Chain-of-Thought (MA-CoT),
the most structured approach to date, achieving a 94.5\% reduction on
LLMSecEval by combining task-specific CWE mitigation guidance, universal
security rules, and language-specific safeguards in a single prompt. The
remediation prompt structure used in this work is adopted from MA-CoT.

\subsection{Hybrid SAST and LLM triage}
Agrawal and Ahi~\cite{agrawal2025} proposed SAST-Genius, a hybrid
framework combining traditional static analysis with an LLM triage layer
that determines whether each finding is a true or false positive based on
full code context and data flow relationships, achieving a 91\% reduction
in false positives. SAST-Genius is the closest existing work to the
pipeline presented here. The key architectural difference is in the
enrichment layer: SAST-Genius enriches with code context, specifically
data flow and taint paths, while this pipeline enriches with threat
context, specifically MITRE ATT\&CK techniques and CWE Observed Examples,
grounding the fix in real-world adversarial behavior rather than static
code structure.

\subsection{Threat modeling and ATT\&CK enrichment}
Elsharef et al.~\cite{elsharef2024} demonstrated that LLMs grounded in NVD
vulnerability data via retrieval-augmented generation can produce
accurate, system-specific threat hypotheses, validating the use of
structured vulnerability knowledge bases as LLM context.
ThreatCompute~\cite{wimbauer2025} showed that modular LLM prompting can
automatically map system components to MITRE ATT\&CK techniques with a
93\% discovery rate against a known benchmark. Both works validate the
core enrichment principle used in this pipeline: connecting code-level
vulnerability findings to system-level threat context improves the quality
of security reasoning. This pipeline extends that principle from threat
modeling into remediation itself, using MITRE ATT\&CK techniques and CWE
Observed Examples to directly ground the fixes an LLM generates, an
application that remains largely unexplored in the automated code
generation and repair literature.

\section{Methodology}
This section describes the research design underlying the evaluation,
including the hypotheses, dataset selection, pipeline architecture,
experimental configurations, and the metrics used to assess outcomes.

\subsection{Research Questions and Hypotheses}
This study addresses two hypotheses. H1: an enriched Code Validator LLM
prompt, returned to the Code Generation LLM alongside the code it
originally generated, improves the security of that code, as measured by a
reduction in static analyzer findings after remediation compared to
baseline. H2: providing the Code Generation LLM with initial static
analyzer findings in addition to enriched Code Validator output further
improves remediation outcomes compared to enrichment context alone.

\subsection{Evaluation Dataset}
The evaluation set has 26 natural language coding prompts drawn from the
LLMSecEval benchmark \cite{tony2023}, covering nine weaknesses from the CWE
Top 25: SQL injection (CWE-89), OS command injection (CWE-78), path
traversal (CWE-22), deserialization of untrusted data (CWE-502), use of
hard-coded credentials (CWE-798), cross-site scripting (CWE-79), improper
input validation (CWE-20), insufficiently protected credentials (CWE-522),
and incorrect permission assignment (CWE-732). These CWEs were selected
for their relevance to Python and their coverage by the static analyzers
used in the study. All LLMSecEval prompts for the selected CWEs were
retained, approximately three per weakness and two for CWE-20, yielding 26
prompts. All prompts targeted Python code generation uniformly across the
set. Table~\ref{tab:prompts} lists the 26 selected prompts by ID and CWE
category. Full prompt text, quality scores, and secure reference
implementations are available in the LLMSecEval repository:
\url{https://github.com/tuhh-softsec/LLMSecEval}, and in the project
\texttt{data\textbackslash llmseceval\textbackslash
LLMSecEval-Prompts\_dataset.csv}.

\begin{table}[H]
\centering
\footnotesize
\setlength{\tabcolsep}{5pt}
\caption{The 26 selected LLMSecEval prompts by prompt ID and CWE.}
\label{tab:prompts}
\begin{tabular}{llll}
\toprule
Prompt ID & CWE & Prompt ID & CWE \\
\midrule
CWE-20\_IIV-1b  & CWE-20  & CWE-732\_IPA-3a & CWE-732 \\
CWE-20\_IIV-2a  & CWE-20  & CWE-78\_INS-1b  & CWE-78  \\
CWE-22\_ILP-1a  & CWE-22  & CWE-78\_INS-2c  & CWE-78  \\
CWE-22\_ILP-2c  & CWE-22  & CWE-78\_INS-3a  & CWE-78  \\
CWE-22\_ILP-3a  & CWE-22  & CWE-79\_INI-1c  & CWE-79  \\
CWE-502\_DUD-1a & CWE-502 & CWE-79\_INI-2a  & CWE-79  \\
CWE-502\_DUD-2b & CWE-502 & CWE-79\_INI-3a  & CWE-79  \\
CWE-502\_DUD-3c & CWE-502 & CWE-798\_UHC-1a & CWE-798 \\
CWE-522\_IPC-1b & CWE-522 & CWE-798\_UHC-2b & CWE-798 \\
CWE-522\_IPC-2a & CWE-522 & CWE-798\_UHC-3a & CWE-798 \\
CWE-522\_IPC-3a & CWE-522 & CWE-89\_SQI-1b  & CWE-89  \\
CWE-732\_IPA-1a & CWE-732 & CWE-89\_SQI-2c  & CWE-89  \\
CWE-732\_IPA-2c & CWE-732 & CWE-89\_SQI-3a  & CWE-89  \\
\bottomrule
\end{tabular}
\end{table}

\subsection{Pipeline Architecture}
\begin{figure}[h]
  \centering
  \includegraphics[width=\columnwidth]{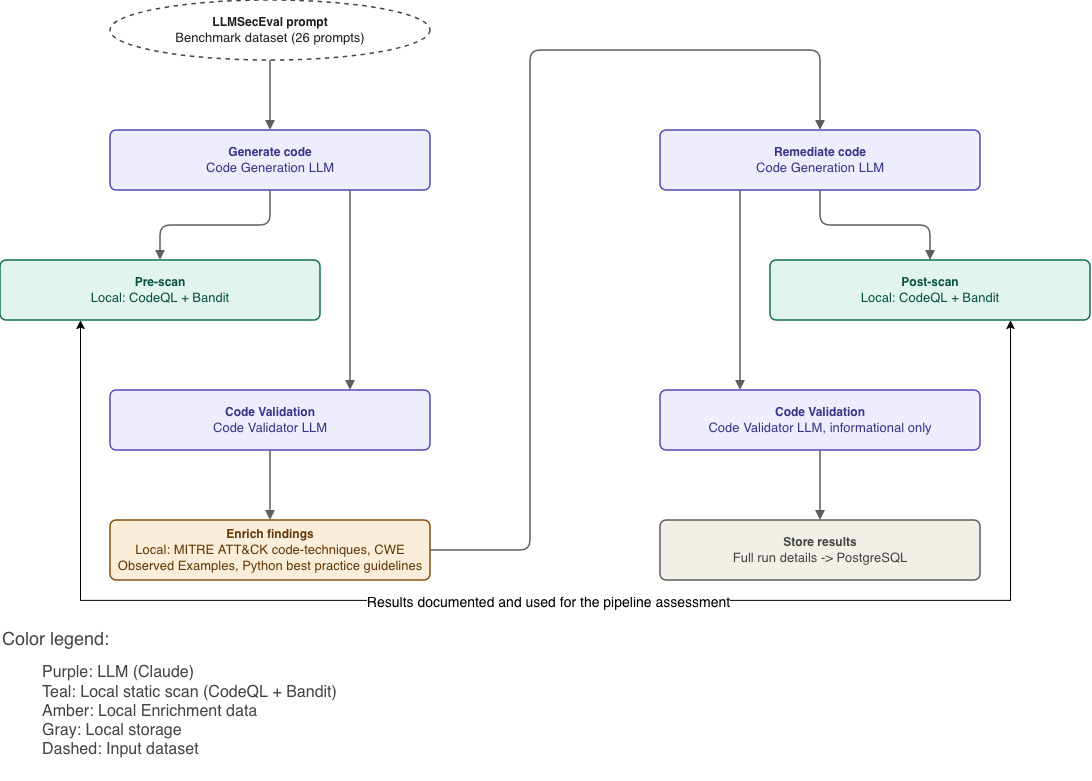}
  \caption{7-stage pipeline workflow with tool labels.}
  \label{fig:pipeline}
  \footnotesize\textit{Note. All figures use static findings
  (CodeQL + Bandit) over complete runs only (26/26 prompts),
  for the four models with ten runs per pipeline. P1 =
  remediation without static-analyzer findings; P2 =
  with static findings.}
\end{figure}

The pipeline is a 7-stage process: code generation, parallel pre-fix
scanning and Code Validator assessment, finding enrichment, code
remediation, post-fix scanning, and result storage. The full pipeline
architecture with tool labels for each stage is illustrated in
Figure~\ref{fig:pipeline}.

\subsection{Experimental Design}
\begin{figure}[h]
  \centering
  \includegraphics[width=\columnwidth]{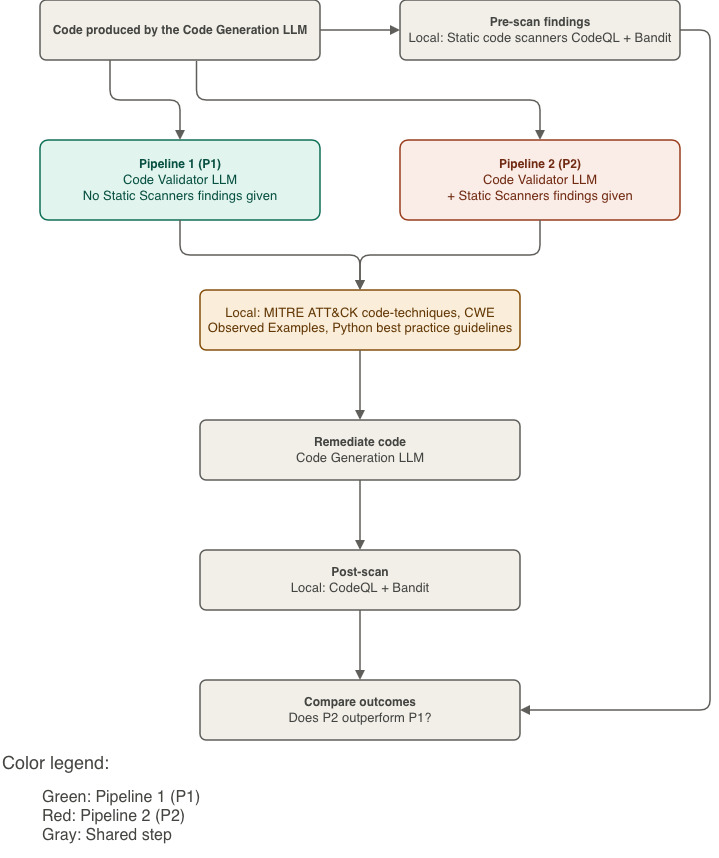}
  \caption{Pipeline 1 vs Pipeline 2 experimental design.}
  \label{fig:pipeline_ab}
\end{figure}

Two pipeline configurations were evaluated. Pipeline 1 (P1) provides the
Code Generation LLM with enriched Code Validator findings and with MITRE
ATT\&CK techniques, CWE Observed Examples, and Python best practice
guidelines only. Pipeline 2 (P2) additionally provides the initial CodeQL
and Bandit findings, testing whether concrete line-specific scanner
evidence improves remediation outcomes beyond enrichment context alone.
Figure~\ref{fig:pipeline_ab} illustrates where the two configurations
diverge and reconverge. Each configuration was run 10 times per model
across four Claude models: Opus 4.8, Sonnet 4.6, Sonnet 5, and Haiku
4.5, producing 80 total runs. Each run covers all 26 prompts, and results
are reported as averages across runs to account for the non-deterministic
nature of LLM output. The two pipeline configurations are illustrated in
Figure~\ref{fig:pipeline_ab}.

\subsection{Baseline and Verdict Logic}
For each prompt, the baseline is defined as the average number of static
analyzer findings (CodeQL and Bandit only) on the freshly generated code
before any remediation, pooled across all runs of a model configuration.
This provides a stable, model-specific reference point independent of the
Code Validator LLM. Verdict logic is count-based: a post-fix static
finding count of zero is a pass, a count lower than baseline is partial,
and a count equal to or higher than baseline is a fail. The Code Validator
LLM runs in both pipelines but its findings do not influence the verdict,
ensuring the evaluation measures objective static analysis outcomes rather
than LLM self-assessment.

\subsection{Metrics}
Four metrics are reported. Resolution rate measures the percentage
reduction in static analyzer findings versus baseline. Verdict
distribution reports the share of pass, partial, and fail outcomes across
all runs per model and configuration. New vulnerability introduction rate
measures the percentage of remediations that introduced at least one
finding type absent from the pre-fix scan. Consistency measures verdict
stability across repeated runs using modal agreement, defined as the
fraction of runs converging on the most common verdict per prompt,
averaged across all 26 prompts.

\section{Implementation / Solution}
The pipeline is implemented in Python 3.12, with PostgreSQL 16 (Docker)
for storage, Streamlit for the UI/monitoring. Rather than using the
Anthropic SDK, all model calls are made by invoking the Claude CLI as a
subprocess, a design chosen for reproducibility and to avoid usage of the
Claude API. Two model roles are configured independently through
environment variables: a Code Generation LLM and a Code Validator LLM,
both recorded per run so results are traceable to the exact models that
produced them.

Static analysis combines CodeQL and Bandit, invoked as subprocesses with
their outputs normalized into a unified finding format (rule ID, mapped
CWE, severity, line) and deduplicated by (CWE, line), with CodeQL treated
as the authoritative source on overlap. Each finding is then enriched
entirely from local, committed data with no network calls at runtime:
MITRE ATT\&CK technique mappings are resolved from an offline dictionary,
CWE Observed Examples are drawn from a pre-parsed MITRE CWE XML file with
up to three CVE references per finding, and per-CWE Python remediation
guidance is embedded in the fix prompt alongside universal security rules.
P1 assembles the fix context from Code Validator findings and enrichment
only; P2 additionally injects a labeled block of the CodeQL and Bandit
pre-fix findings.

The seven pipeline stages execute sequentially per prompt, with scanning
and Code Validator assessment run concurrently via asyncio. Every
subprocess call has an explicit timeout, generated and fixed code is
validated with \texttt{ast.parse} with a single retry on failure, and
every module returns a typed safe-default error result rather than
raising. Security best practices are applied throughout the codebase. The
ongoing details for each step are saved into the database for further
analysis.

A Streamlit dashboard provides five tabs covering live run monitoring,
historical results, consistency analysis, baseline comparison, and
aggregate evaluation findings. Results are exportable as CSV.

Full implementation details including an AES-256 encrypted database
snapshot are available in the project repository (Appendix~\ref{app:repo}).

\section{Evaluation}
This section reports quantitative findings across four dimensions:
hypothesis validation, model rankings, new vulnerability introduction, and
verdict consistency. All results are based on complete runs of 26 prompts
and exclude error verdicts.

\subsection{Hypothesis 1: Does enriched Code Validator output improve
remediation quality (P1)?}
P1 reduced static analyzer findings versus baseline for every model
tested, confirming H1. Reductions ranged from $-9\%$ (Opus 4.8) to
$-54\%$ (Sonnet 5).

\begin{table}[H]
\centering
\caption{Hypothesis 1: baseline versus P1 static analyzer findings.}
\label{tab:h1}
\begin{tabular}{lccc}
\toprule
Model & Baseline & After P1 & Reduction \\
\midrule
Haiku 4.5  & 2.03 & 1.04 & $-49\%$ \\
Opus 4.8   & 0.82 & 0.75 & $-9\%$  \\
Sonnet 4.6 & 1.09 & 0.66 & $-39\%$ \\
Sonnet 5   & 1.44 & 0.67 & $-54\%$ \\
\midrule
Pooled     & 1.34 & 0.78 & $-42\%$ \\
\bottomrule
\end{tabular}
\end{table}

Opus 4.8 shows the smallest improvement not because remediation is less
effective, but because its baseline is already the cleanest of the four
models.

\subsection{Hypothesis 2: Does adding static analyzer findings further
improve outcomes (P2)?}
P2 outperformed P1 for every model, confirming H2. The pooled average
reduction deepened from $-42\%$ under P1 to $-57\%$ under P2, with the
direction unanimous across all four models.

\begin{table}[H]
\centering
\caption{Hypothesis 2: P1 and P2 reductions versus baseline.}
\label{tab:h2}
\begin{tabular}{lcccc}
\toprule
Model & Baseline & P1 vs base & P2 vs base & P2 vs P1 \\
\midrule
Haiku 4.5  & 2.03 & $-49\%$ & $-69\%$ & $-40\%$ \\
Opus 4.8   & 0.82 & $-9\%$  & $-29\%$ & $-23\%$ \\
Sonnet 4.6 & 1.09 & $-39\%$ & $-56\%$ & $-27\%$ \\
Sonnet 5   & 1.44 & $-54\%$ & $-56\%$ & $-4\%$  \\
\midrule
Pooled     & 1.34 & $-42\%$ & $-57\%$ & $-26\%$ \\
\bottomrule
\end{tabular}
\end{table}

\subsection{Model Rankings}
The best code generator is not the best pipeline performer. Opus 4.8
produces the most secure first draft (baseline 0.82, 54\% clean at
generation). Table~\ref{tab:rankings_baseline} shows the full baseline
rankings across all four models. Sonnet 4.6 produces the best outcome
after P2 remediation (lowest residual 0.48, highest pass rate 75.4\%).
Table~\ref{tab:rankings_p2} shows the rankings after P2 remediation. The
P2 pipeline compresses the performance gap between models: baselines
ranged from 0.82 to 2.03 across models, while post-P2 residuals converged
to a tight band of 0.48 to 0.64.

\begin{table}[h]
\centering
\caption{Model rankings: baseline code quality.}
\label{tab:rankings_baseline}
\begin{tabular}{clcc}
\toprule
Rank & Model & Baseline avg findings & Clean at generation \\
\midrule
1 & Opus 4.8    & 0.82 & 54.4\% \\
2 & Sonnet 4.6  & 1.09 & 46.9\% \\
3 & Sonnet 5    & 1.44 & 34.8\% \\
4 & Haiku 4.5   & 2.03 & 30.6\% \\
\bottomrule
\end{tabular}
\end{table}

\begin{table}[h]
\centering
\caption{Model rankings: after P2 remediation.}
\label{tab:rankings_p2}
\begin{tabular}{clcc}
\toprule
Rank & Model & P2 residual & Pass rate after P2 \\
\midrule
1 & Sonnet 4.6  & 0.48 & 75.4\% \\
2 & Opus 4.8    & 0.58 & 69.6\% \\
3 & Haiku 4.5   & 0.62 & 67.3\% \\
4 & Sonnet 5    & 0.64 & 65.4\% \\
\bottomrule
\end{tabular}
\end{table}

\subsection{New Vulnerability Introduction}
Remediation occasionally introduced new weaknesses. Across all models and
configurations, 15--22\% of remediations introduced at least one new
finding. P2 reduced this churn for three of four models; Sonnet 5 was the
sole exception, where churn increased from 20.4\% under P1 to 21.9\% under
P2. The full breakdown per model and configuration is shown in
Table~\ref{tab:newvulns_full}.

\begin{table}[H]
\centering
\caption{Percentage of remediations introducing at least one new finding.}
\label{tab:newvuln}
\begin{tabular}{lcc}
\toprule
Model & P1 \% with new & P2 \% with new \\
\midrule
Haiku 4.5  & 21.5\% & 17.7\% \\
Opus 4.8   & 18.5\% & 18.1\% \\
Sonnet 4.6 & 18.5\% & 15.0\% \\
Sonnet 5   & 20.4\% & 21.9\% \\
\bottomrule
\end{tabular}
\end{table}

\begin{table}[H]
\centering
\caption{Newly-introduced vulnerabilities: full breakdown.}
\label{tab:newvulns_full}
\begin{tabular}{llcccc}
\toprule
Model & P & \% $\geq$1 & Total & Avg & 1 / 2 / 3+ \\
\midrule
Opus 4.8    & P1 & 18.5\% & 67 & 1.40 & 38 / 5 / 5 \\
Opus 4.8    & P2 & 18.1\% & 59 & 1.26 & 38 / 7 / 2 \\
Sonnet 4.6  & P1 & 18.5\% & 64 & 1.33 & 38 / 4 / 6 \\
Sonnet 4.6  & P2 & 15.0\% & 57 & 1.46 & 27 / 6 / 6 \\
Sonnet 5    & P1 & 20.4\% & 82 & 1.55 & 35 / 9 / 9 \\
Sonnet 5    & P2 & 21.9\% & 96 & 1.68 & 31 / 14 / 12 \\
Haiku 4.5   & P1 & 21.5\% & 82 & 1.46 & 36 / 14 / 6 \\
Haiku 4.5   & P2 & 17.7\% & 66 & 1.43 & 31 / 10 / 5 \\
\midrule
Pooled      & P1 & 19.7\% & 295 & 1.44 & 147 / 32 / 26 \\
Pooled      & P2 & 18.2\% & 278 & 1.47 & 127 / 37 / 25 \\
\bottomrule
\end{tabular}
\end{table}

\subsection{Verdict Consistency}
Verdict stability averaged approximately 81\% modal agreement across all
models and configurations, indicating that a given prompt reached the same
verdict in roughly four of every five runs. P2 was marginally more stable
than P1. The full per-model and per-configuration breakdown is shown in
Table~\ref{tab:consistency}. Full per-model and per-configuration
consistency breakdowns are available in the project repository.

\begin{table}[h]
\caption{Verdict consistency: modal agreement across 10 runs.}
\label{tab:consistency}
\begin{tabular}{llcc}
\toprule
Model & Pipeline & Avg agreement & Fully consistent (of 26) \\
\midrule
Haiku 4.5   & P1 & 70.4\% & 4 \\
Haiku 4.5   & P2 & 78.1\% & 6 \\
Opus 4.8    & P1 & 83.1\% & 7 \\
Opus 4.8    & P2 & 82.7\% & 11 \\
Sonnet 4.6  & P1 & 81.9\% & 8 \\
Sonnet 4.6  & P2 & 85.8\% & 10 \\
Sonnet 5    & P1 & 79.6\% & 9 \\
Sonnet 5    & P2 & 83.8\% & 8 \\
\midrule
Pooled      & P1      & 78.8\% & 7.0 \\
Pooled      & P2      & 82.6\% & 8.8 \\
Pooled      & P1 + P2 & 80.7\% & 7.9 \\
\bottomrule
\end{tabular}
\end{table}

\begin{quote}
\footnotesize\textit{Modal agreement = per prompt, the share
of runs landing on the most common verdict, averaged over the
26 prompts. Reported as a secondary observation confirming
run-to-run variability was tracked.}
\end{quote}

\section{Discussion \& Analysis}
Both hypotheses were confirmed across all four models and in a unanimous
direction. P1 demonstrated that returning enriched Code Validator findings
to the Code Generation LLM consistently reduces static analyzer findings
versus baseline, and P2 demonstrated that adding initial static analyzer
findings to that context deepens the reduction further. The unanimity
across models of different sizes and capabilities is notable: the
enrichment approach is not model-specific, it generalizes.

Our results are consistent with the direction established by prior work on
CWE-specific prompting. Bruni et al.~\cite{bruni2025} reported reductions
of 56--68.7\% using prompt engineering techniques on GPT models; Aldosari
and Aldawsari~\cite{aldosari2026} achieved up to 77\% with CWE-specific
meta-prompting across open-source models; Kharma et al.~\cite{kharma2026}
reached 94.5\% on LLMSecEval using MA-CoT. Our P1 reduction of 42\% and P2
reduction of 57\% fall within or near this range. Direct numerical
comparison is not appropriate given the differences in models, languages,
CWE coverage, prompt structure, and evaluation setup across these studies.
What the literature establishes, and what this work confirms, is that
structured, CWE-aware context consistently improves remediation outcomes.

A notable finding is that the best code generator is not the best pipeline
performer. Opus 4.8 produces the most secure first draft yet finishes
second after P2 remediation, while Sonnet 4.6 starts with more findings
and ends the cleanest. This suggests that generation quality and
remediation effectiveness are distinct properties, and that selecting a
model purely on baseline security may not optimize end-to-end pipeline
outcomes. Practitioners building remediation pipelines should evaluate
models on post-remediation outcomes, not pre-remediation baselines.

Remediation introduced new vulnerabilities in 15--22\% of cases across all
models and configurations, confirming that automated fixing carries an
inherent regression risk. P2 reduced this churn for three of four models,
suggesting that richer fix context not only improves resolution but also
reduces the likelihood of introducing new issues. This consistent
regression risk across all models and configurations underscores the
necessity of post-fix verification as a first-class component of any
automated remediation pipeline.

\section{Limitations}
This study is scoped to Python code generation across nine CWE categories
and 26 prompts, using four Claude models from the same model family.
Findings may not generalize to other programming languages, broader CWE
coverage, or models from different providers or architectures.

Security evaluation relies exclusively on static analysis (\mbox{CodeQL} and
Bandit). These tools detect known vulnerability patterns but miss logic
errors, authentication flaws, race conditions, and context-dependent
vulnerabilities that only manifest at runtime. A passing verdict indicates
no detected patterns, not confirmed security. Additionally, vulnerability
detection does not equal exploitability: a flagged finding may not be
reachable or exploitable in a real deployment context.

The verdict logic and resolution metrics treat all static analyzer
findings equally, regardless of severity. A high-severity SQL injection
finding and a low-severity informational warning both count as one finding
toward the verdict. This means the pipeline does not distinguish between
remediations that eliminated critical vulnerabilities and those that
resolved minor issues, potentially overstating or understating the
practical security impact of the remediation. Future work should
incorporate severity-weighted metrics to provide a more nuanced assessment
of pipeline effectiveness.

Two further constraints affect the generalizability of results. First,
Claude models may have been exposed to LLMSecEval prompts during training,
which could inflate improvement rates relative to a truly unseen
benchmark. Second, the study does not evaluate functional correctness:
security improvements in the fixed code could introduce functionality
regressions that are not captured by the evaluation metrics. Finally, with
10 runs per configuration, results represent statistical tendencies rather
than guarantees, and the pooled baseline design is one approach to
establishing a reference point.

\section{Threats to Validity}
\textit{Internal validity.} The primary internal threat is that the
verdict logic is count-based rather than exploitability-based: a reduction
in static findings is treated as a security improvement, but this does not
confirm that the remaining findings are unexploitable or that the resolved
findings were genuinely dangerous. A secondary threat is that the Code
Validator LLM drives the remediation prompt in P1, while in P2 it is
supplemented with the static findings: if the validator LLM misclassifies
or misses a vulnerability, the fix is built on incomplete input. Finally,
LLM output is stochastic, and while 10 runs per configuration mitigates
this, individual run results may not be fully representative of the
underlying distribution.

\textit{External validity.} The study is limited to Python, nine CWE
categories, 26 synthetic prompts from LLMSecEval, and four Claude models
from the same model family. Results may not generalize to other languages,
a broader set of weaknesses, real-world developer code, or models from
different providers. The synthetic nature of LLMSecEval prompts means the
vulnerability patterns elicited may differ from those that arise
organically in production codebases.

\section{Future Work}
The most immediate extension is expanding the scope of evaluation. The
current study is limited to Python and nine CWE categories; future work
should evaluate the pipeline across additional languages such as
JavaScript and Java, and against a broader set of weaknesses from the
MITRE CWE Top 25. Testing across models from different providers and
architectures, including open-source models, would strengthen the
generalizability of the findings.

The evaluation methodology could also be strengthened. The current study
measures security outcomes only; future work should incorporate functional
correctness testing to verify that fixed code remains executable and
behaviorally correct. Complementing static analysis with runtime and
dynamic testing would capture vulnerability classes that CodeQL and Bandit
cannot detect.

Incorporating severity-weighted metrics into the verdict logic would
provide a more nuanced assessment of pipeline effectiveness,
distinguishing remediations that eliminate critical vulnerabilities from
those that resolve minor findings. A CWE-level breakdown of remediation
outcomes would provide additional granularity, revealing which
vulnerability categories respond most and least effectively to the
enrichment-based remediation approach.

Two pipeline extensions are worth exploring. First, integrating the
pipeline into developer workflows as a real-time tool, for example as an
IDE plugin or a CI/CD gate would test whether the security improvements
observed in batch evaluation translate to practical adoption. Second, the
current design pairs the same model in both the Code Generation and Code
Validator roles within each configuration. Testing cross-model pairings
for example using a stronger model as the Code Validator with a smaller
model as the Code Generation LLM could reveal whether validator quality or
generator quality is the more important factor in determining pipeline
outcomes.

\section{Summary and Conclusion}
AI-assisted development tools introduce security vulnerabilities at
significant rates, yet automated mechanisms that ground remediation in
real-world threat context remain rare. This paper presented an automated
security evaluation pipeline built around two LLM roles, a Code Generation
LLM and a Code Validator LLM, that enriches findings with MITRE ATT\&CK
techniques and CWE Observed Examples and uses that enrichment directly in
the remediation prompt, alongside standard Python-specific best practices
and post-fix static verification. Two configurations were evaluated:
Pipeline 1, using enriched Code Validator output only, and Pipeline 2,
additionally providing static analyzer findings to the fix context.

Both hypotheses were confirmed across all four models and in a unanimous
direction, as detailed in Section 6. The enriched Code Validator approach
consistently reduced static analyzer findings versus baseline, and adding
static analyzer findings to the fix context deepened those reductions
further for every model. A notable secondary finding is that the best code
generator is not the best pipeline performer, suggesting that generation
quality and remediation effectiveness are distinct properties that should
be evaluated independently.

The results establish that structured, CWE-aware enrichment generalizes
across models of different sizes and capabilities, and that providing
richer fix context, including concrete scanner evidence, consistently
improves outcomes. The new vulnerability introduction rates observed
across all configurations confirm that automated remediation carries an
inherent regression risk, validating post-fix verification as a necessary
component of any automated security pipeline rather than an optional step.

\section{AI Usage Acknowledgement}
Claude (claude.ai) was used for grammar checking of written work;
corrections were limited to grammar and spelling, with the original
language and style preserved. Claude Code was used as a coding co-pilot
throughout the implementation phase, providing architectural guidance,
writing and debugging code, under the author's direction and review. All
research conclusions, design decisions, and written content are the
author's own. The full project codebase, including the CLAUDE.md
specification file that guided Claude Code throughout the build, is
available in the project repository (Appendix~\ref{app:repo}).

\bibliographystyle{ACM-Reference-Format}
\bibliography{references}

@article{aldosari2026,
  author = {Aldosari, Shaykhah S. and Aldawsari, Layla S.},
  title = {Securing LLM Code Generation: Leveraging Prompt Engineering
           to Mitigate Vulnerabilities Across Models and Languages},
  journal = {Science of Computer Programming},
  volume = {251},
  pages = {103446},
  year = {2026},
  doi = {10.1016/j.scico.2026.103446}
}

@misc{agrawal2025,
  author = {Agrawal, Vaibhav and Ahi, Kiarash},
  title = {{LLM}-Driven {SAST}-Genius: A Hybrid Static Analysis
           Framework for Comprehensive and Actionable Security},
  year = {2025},
  eprint = {2509.15433},
  archivePrefix = {arXiv}
}

@misc{bruni2025,
  author = {Bruni, Marc and Gabrielli, Fabio and Ghafari, Mohammad
            and Kropp, Martin},
  title = {Benchmarking Prompt Engineering for Secure Code Generation},
  year = {2025},
  eprint = {2502.06039},
  archivePrefix = {arXiv}
}

@inproceedings{elsharef2024,
  author = {Elsharef, Isra and Zeng, Zhen and Gu, Zhongshu},
  title = {Facilitating Threat Modeling by Leveraging Large Language Models},
  booktitle = {Proceedings of the Workshop on AI Systems with
               Confidential Computing (AISCC), NDSS Symposium},
  year = {2024},
  url = {https://www.ndss-symposium.org/ndss-paper/auto-draft-539/}
}

@techreport{ji2024,
  author = {Ji, Jummy and Jun, Janet and Wu, Micah and Gelles, Rebecca},
  title = {Cybersecurity Risks of {AI}-Generated Code},
  institution = {Center for Security and Emerging Technology,
                 Georgetown University},
  year = {2024},
  url = {https://cset.georgetown.edu/publication/cybersecurity-risks-of-ai-generated-code/}
}

@misc{kharma2026,
  author = {Kharma, Mohammed F. and Alkhanafseh, Mohammad and
            Sabbah, Ahmed and Mohaisen, David},
  title = {Enhancing Reliability in {LLM}-Based Secure Code Generation},
  year = {2026},
  eprint = {2605.24300},
  archivePrefix = {arXiv}
}

@misc{mitre2023,
  author = {{MITRE}},
  title = {{CWE} Top 25 Most Dangerous Software Weaknesses},
  year = {2023},
  url = {https://cwe.mitre.org/top25/}
}

@misc{mitre2024,
  author = {{MITRE}},
  title = {{MITRE ATT\&CK} Framework},
  year = {2024},
  url = {https://attack.mitre.org/}
}

@misc{schreiber2025,
  author = {Schreiber, Maximilian and Tippe, Pascal},
  title = {Security Vulnerabilities in {AI}-Generated Code:
           A Large-Scale Analysis of Public {GitHub} Repositories},
  year = {2025},
  eprint = {2510.26103},
  archivePrefix = {arXiv}
}

@inproceedings{tony2023,
  author = {Tony, Catherine and Mutas, Markus and
            D{\'{i}}az Ferreyra, Nicol{\'{a}}s E. and
            Scandariato, Riccardo},
  title = {{LLMSecEval}: A Dataset of Natural Language Prompts
           for Security Evaluations},
  booktitle = {Proceedings of the 20th IEEE/ACM International
               Conference on Mining Software Repositories (MSR)},
  year = {2023},
  doi = {10.5281/zenodo.7565965}
}

@misc{wessling2025,
  author = {Wessling, Jens},
  title = {We Asked 100+ {AI} Models to Write Code.
           Here's How Many Failed Security Tests},
  year = {2025},
  month = {July},
  url = {https://www.veracode.com/blog/genai-code-security-report/}
}

@inproceedings{wimbauer2025,
  author = {Wimbauer, Anna and Muscariello, Luca and Samain, Jacques
            and Steger, Lion and Glas, Kilian and Helm, Max
            and Carle, Georg},
  title = {{ThreatCompute}: Leveraging {LLMs} for Automated Threat
           Modeling of Cloud-Native Applications},
  booktitle = {Proceedings of the 2025 Cloud Computing Security
               Workshop (CCSW '25)},
  year = {2025},
  doi = {10.1145/3733812.3765533}
}

\appendix

\section{Project Repository}
\label{app:repo}
Full project codebase, and AES-256 encrypted database snapshot:
\url{https://github.gatech.edu/msurikov3/Securing_AT_Generated_Code}

\noindent A public mirror of the full project is also available at:
\url{https://github.com/Droidsurikov/securing-ai-generated-code}

\end{document}